\documentstyle[prl,aps,floats,psfig]{revtex}
\catcode`@=11

\draft

\def\vev#1{{\langle#1\rangle}}

\def\lsim{\mathrel{\raise.3ex\hbox{$<$\kern-.75em\lower1ex\hbox{$\sim$}}}}
\def\gsim{\mathrel{\raise.3ex\hbox{$>$\kern-.75em\lower1ex\hbox{$\sim$}}}}

\begin{document}

\twocolumn[\hsize\textwidth\columnwidth\hsize\csname
@twocolumnfalse\endcsname

\hfill\vbox{
\hbox{BUHEP-01-34}
\hbox{MADPH-01-1252}
\hbox{hep-ph/0112125}
\hbox{}}

\title{Inverting a Supernova: Neutrino Mixing, Temperatures and Binding Energy}
\author{$^1$V. Barger, $^2$D. Marfatia and $^1$B. P. Wood}
\vskip 0.3in
\address{$^1$Department of Physics, University of Wisconsin, Madison, WI 53706}
\vskip 0.1in
\address{$^2$Department of Physics, Boston University, Boston, MA 02215}
\maketitle

\begin{abstract}

We show that 
the temperatures of the emergent non-electron neutrinos and the 
binding energy 
released by a galactic Type II supernova are determinable, assuming the
Large Mixing Angle (LMA) solution is correct,
 from observations at the Sudbury Neutrino Observatory (SNO) and at 
Super-Kamiokande (SK). 
 If the neutrino mass hierarchy is inverted, either a lower or upper bound can
be placed on the neutrino mixing angle $\theta_{13}$, 
and the hierarchy can be deduced for
adiabatic transitions. For the normal hierarchy, neither can 
$\theta_{13}$ be constrained nor can the hierarchy be determined. 
Our
conclusions are qualitatively unchanged for the proposed Hyper-Kamiokande
detector.

\end{abstract}
]
Neutrino oscillations convincingly explain the solar and atmospheric neutrino 
anomalies~\cite{solar,atm}. 
Atmospheric data and initial K2K data~\cite{k2k} indicate  
oscillations with
$|\Delta m^2_{31}\equiv m_3^2-m_1^2| \sim 3\times 10^{-3}$ eV$^2$ and
 \mbox{$\sin^2 2\theta_{23} \sim 1$}~\cite{kamk2k}
that will be tested to within 10\% accuracy at MINOS~\cite{minos,mbl}.
(For the standard 
parameterization of the neutrino mixing matrix see Ref.~\cite{pdg}). 
The large mixing angle (LMA) solution 
($\Delta m^2_{21} \sim 5\times 10^{-5}$ eV$^2$, $\sin^2 2\theta_{12} \sim 0.8$)
which is emerging as the
solution to the solar neutrino problem~\cite{global} will be
tested to 10\% accuracy~\cite{kland1} at KamLAND~\cite{kamland}.
Thus, in the near future, all parameters relevant to neutrino oscillations
will be known, except $\sin^22\theta_{13}$, 
which is bounded above by 0.1 at the 95\%
C.L. by the CHOOZ experiment~\cite{chooz}, the sign of 
$\Delta m^2_{31}$ and the CP violating phase. 
In the longer term, long baseline neutrino experiments
using upgraded conventional neutrino beams could achieve a sensitivity
to $\sin^22\theta_{13}$ of about $10^{-3}$~\cite{super}, but neutrinos from
a galactic supernova can probe values that are more than 
two orders of magnitude smaller.

The objective of this Letter is to determine what the
 expected neutrino signals at SNO and SK from  
a Type II galactic supernova can tell us about 
$\sin^22\theta_{13}$, sgn($\Delta m^2_{31}$),
the neutrino 
temperatures, and the binding energy released in such an event 
if the LMA solution is confirmed. Throughout, we assume that solar
and atmospheric parameters will be known to within 10\% from
upcoming experiments. Since low energy $\nu_{\mu}$ and $\nu_\tau$ are 
indistinguishable at SNO and SK, only their transitions
with $\nu_e$ can be studied.
 
\vskip 0.1in
\noindent
\underline{Supernova neutrinos}:
During the early stages of a supernova explosion, as the shock wave rebounds
from the dense inner core of the star and crosses 
the electron neutrinosphere, $\nu_e$'s from electron
capture on protons are released resulting in a breakout or neutronization
burst that carries away $\sim 10^{51}$ ergs. The duration of this burst lasts
only a few milliseconds (no more than 10) and any non-electron neutrino
events at SNO during this time are a consequence of 
$\nu_e\rightarrow \nu_{\mu,\tau}$ oscillations. 
For progenitor stars of mass $\sim 15 M_{\odot}$, 
numerical simulations find that following the neutronization burst,  
99\% of the binding energy released, 
$E_b=1.5-4.5 \times 10^{53}$ ergs, is roughly equipartitioned
 in the form of neutrinos and antineutrinos of all flavors~\cite{mod}. 
Including effects of nucleon bremsstrahlung and 
electron neutrino pair annihilation, the
luminosities are approximately related by 
$L_{\nu_e} \sim L_{\bar{\nu}_e} \sim (1-2)\, L_{\nu_x}$ where 
\mbox{$x = \mu, \bar{\mu}, \tau, \bar{\tau}$}~\cite{lowtau,violation}.
This emission occurs on
a timescale of tens of seconds. 
The mean energies of the different flavors of neutrinos are determined by
 the strength of their interactions with matter, with the most strongly
interacting neutrinos leaving the star with the lowest mean energy 
{\it i.e.}, $\langle E_{\nu_e} \rangle < \langle E_{\bar{\nu}_e} \rangle <
\langle E_{{\nu}_x} \rangle$. The 
authors of Ref.~\cite{lowtau} (see also Ref.~\cite{pinto}) 
emphasize that spectral differences are very small,
 typically
$\langle E_{\nu_e} \rangle : \langle E_{\bar{\nu}_e} \rangle :
\langle E_{{\nu}_x} \rangle :: 0.85 : 1 : 1.1$. The spectra of neutrinos can
be modeled by pinched Fermi-Dirac distributions. We can write the 
unoscillated 
differential flux at a distance $D$ from the 
supernova as
\begin{equation}
F_{\alpha} = \frac{L_{\alpha}}{24\pi D^2 \, T^4_{\alpha} |Li_4(-e^{\eta_{\alpha}})|} \: \frac{E^2}{e^{E/T_{\alpha}-\eta_\alpha} \, + \, 1} \:,
\label{fermi}
\end{equation}
where $\alpha=\nu_e, \bar{\nu}_e, \nu_x$, $Li_n(z)$ is the 
polylogarithm function and $\eta_{\alpha}$ is the 
degeneracy parameter.  
The temperature of the neutrinos, $T_{\alpha}$, is related to 
$\vev{E_{\alpha}}$ via $\vev{E_{\alpha}}=
3 {Li_4(-e^{\eta_{\alpha}}) \over Li_3(-e^{\eta_{\alpha}})}T_{\alpha}$.
We shall use $\vev{E_{\alpha}}$ and $T_{\alpha}$ interchangeably since they
are equivalent to each other once $\eta_{\alpha}$ is specified.
Strictly speaking, weak magnetism
effects may result in $T_{\bar{\nu}_{\mu,\tau}}$ being about 7\% higher than 
$T_{\nu_{\mu,\tau}}$~\cite{horowitz}. However, we have explicitly
checked that the inequality of these temperatures does not affect our 
results. 

As the neutrinos leave the star, they encounter a density profile that 
falls like $1/r^3$~\cite{dense}. If the mass hierarchy is 
normal, {\it i.e.} $\Delta m^2_{31}>0$, (inverted,  {\it i.e.} 
$\Delta m^2_{31}<0$), 
neutrinos (antineutrinos) pass 
through a resonance at high densities 
($10^3-10^4$ g/cm$^3$) which is characterized
by ($\Delta m^2_{31}, \sin^2 2\theta_{13}$) and the neutrinos pass through a
 second resonance at low densities
($\sim 20$ g/cm$^3$ for the LMA solution) that is determined by 
($\Delta m^2_{21}, \sin^2 2\theta_{12}$)~\cite{dighe}. 
Transitions in the latter
resonance are almost adiabatic, with an essentially zero probability of
level crossing. We denote the jumping
probability in the high density resonance by $P_H$ and adopt the potential,
$V_0 (R_{\odot}/r)^3$ with $V_0=1.25 \times 10^{-14}$ eV and the solar radius,
$R_{\odot}=6.96\times 10^{10}$ cm.
 Note that 
$P_H$ is the same for both neutrinos and antineutrinos~\cite{fogli}
and has an 
$e^{-\sin^2 \theta_{13} (|\Delta m^2_{31}|/E_{\nu})^{2/3} V_0^{1/3} }$
dependence~\cite{kuo}. Thus, even an order of magnitude uncertainty in $V_0$
does not have a qualitatively significant effect on $P_H$.

\vskip 0.1in
\noindent
\underline{The integrated spectra at SNO and SK:}
Information on the neutrinos emerging from the supernova after the 
neutronization burst will be contained in 
$\nu_e$ and $\bar{\nu}_e$ spectra observed at SNO and SK. 

For the normal hierarchy, 
the $\nu_e$ flux will be partially 
or completely converted into $\nu_{\mu}$ and 
$\nu_{\tau}$ with the survival probability given by~\cite{dighe}
\begin{equation}
P=P_HP_{2e} + (1-P_H)\sin^2\theta_{13}  \label{prob}\,.
\end{equation}
The sensitivity of the signal depends on 
$\sin^22\theta_{13}$ both explicitly
and implicitly \mbox{through $P_H$}. 
The survival probability for electron antineutrinos 
is $\bar{P}=\bar{P}_{1e}$~\cite{dighe}, which
is the probability that an antineutrino reaching the earth
in the $\bar{\nu}_1$ mass eigenstate interacts as a $\bar{\nu}_e$.

In the case of the inverted hierarchy, the $\nu_e$ survival probability is
$P=P_{2e}$~\cite{dighe}, which is the probability that a neutrino reaching 
the earth in the $\nu_2$ mass eigenstate 
will interact in a detector as $\nu_e$.
The $\bar{\nu}_e$ survival probability is~\cite{dighe}
\begin{equation}
\bar{P}=P_H \bar{P}_{1e}+(1-{P}_H)\sin^2 \theta_{13}\,.
\end{equation}
Since $\bar{P}_{1e}$ and $P_{2e}$ depend only on oscillation parameters at 
the solar scale (and the supernova's zenith angle $\theta_Z$),
nothing can be learned about  $\sin^22\theta_{13}$ from the $\bar{\nu}_e$ ($\nu_e$)
 flux if the hierarchy is normal (inverted).  
\begin{figure}[t]
\mbox{\psfig{file=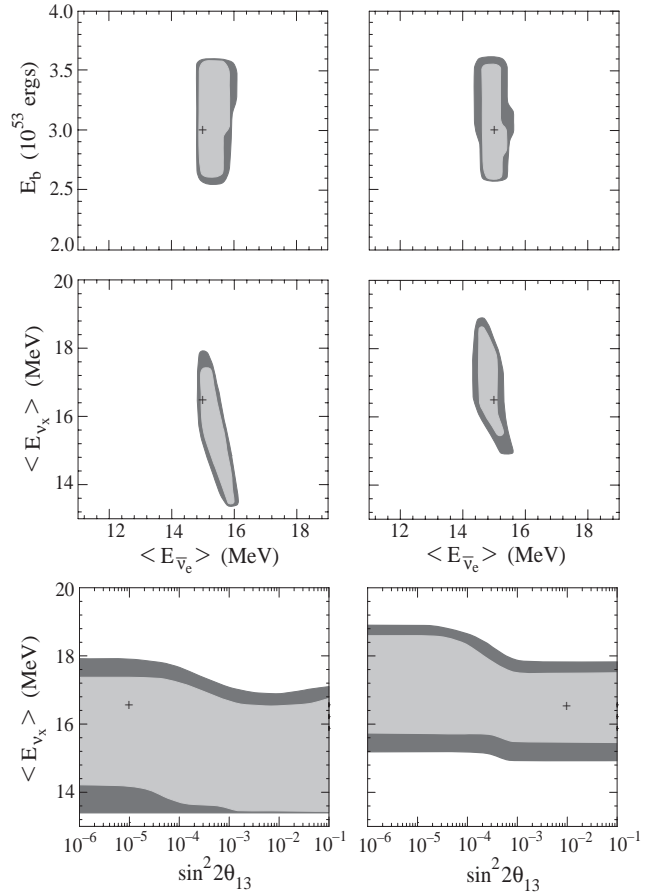,width=8.5cm,height=12cm}}
\medskip
\caption{Determination of 
 the binding energy $E_b$, the supernova
neutrino mean energies (temperatures) and $\sin^22\theta_{13}$ 
for the normal mass hierarchy. The left-hand
and right-hand panels correspond to data simulated at 
$\sin^22\theta_{13} = 10^{-5}$ ($P_H=1$)
and $\sin^22\theta_{13} = 10^{-2}$ ($P_H=0$), respectively.
The cross-hairs 
mark the theoretical inputs, and the $90$\% and 
$99\%$ C.L. regions are light and dark shadings, respectively. 
$\vev{E_{{\nu}_x}}$ is
the mean energy of the non-electron neutrinos.}
\label{fig:1reg}
\end{figure}

For either hierarchy, we expect little sensitivity to $T_{\nu_e}$ because
the survival probability of electron neutrinos is no
more than about $P_{2e}\sim \sin^2 \theta_{12} \sim 0.3$. 

For the 32 kton fiducial volume of the reinstrumented
SK detector (with a 7.5 MeV threshold), 
and the 1.4 kton fiducial volume of the light water tank at SNO 
(with a threshold of 5 MeV), we only consider events
that are isotropic and indistinguishable from each other; \mbox{they are}
\begin{eqnarray}
\bar{\nu}_e + p\ \,  &&\rightarrow n + e^+ \label{nep}\,,\\
\nu_e + O \rightarrow F + e^-\,,&&\ \ \ 
\bar{\nu}_e + O \rightarrow N + e^+ \label{neo}\,.
\end{eqnarray}
We do not consider electron scattering events. 
The good directional capability on these events
allows their separation, and they play an important role
in the reconstruction of the direction of the supernova that in turn determines
the extent to which earth matter effects may be important~\cite{lunardini}.

Neutrinos will interact with deuterium in the 1 kton 
fiducial volume of the
heavy water tank at SNO (with a 5 MeV threshold) 
via the charged current $(CC)$
reactions,
\begin{eqnarray}
\nu_e + d &\rightarrow& p + p + e^- \label{snod}\,,\\
\bar{\nu}_e + d &\rightarrow& n+n + e^+ \label{nebard}\,.
%\nu_y + d &\rightarrow& \nu_y + p + n\,,\ \ \ \ \ \ \  y=e,\mu,\tau \label{snoxd}\,.
\end{eqnarray}
In addition we include the reactions of Eq.~(\ref{neo}).
Two neutron captures in addition to a Cherenkov light cone can distinguish 
$\bar{\nu}_e$-$d$ events from the other 
charged current scattering events on deuterium or oxygen.
All $NC$ events
(whose signal is a single neutron capture and no electron), and
electron scattering events are neglected.

\begin{figure}[t]
\mbox{\psfig{file=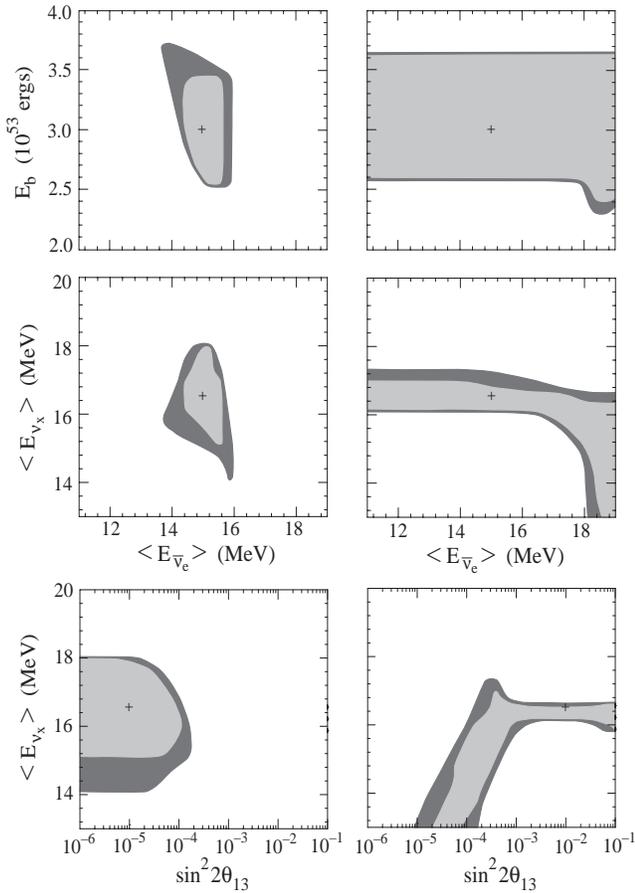,width=8.5cm,height=12cm}}
\medskip
\caption{The same as Fig.~\ref{fig:1reg} but for the inverted hierarchy.}
\label{fig:inv}
\end{figure}

To simulate the energy
spectra for the channels under consideration at the two experiments we assume a 
typical supernova~\cite{lowtau} at a distance of 10 kpc: 
$E_b = 3 \times 10^{53}$ ergs, 
$\langle E_{\nu_e} \rangle : \langle E_{\bar{\nu}_e} \rangle :
\langle E_{{\nu}_x} \rangle :: 0.85 : 1 : 1.1$ with 
$\langle E_{\bar{\nu}_e} \rangle = 15$ MeV, $\eta_{\nu_e}=2$ 
$(T_{\nu_e}=  \langle E_{\nu_e} \rangle/3.61) $, 
$\eta_{\bar{\nu}_e}=3$ $(T_{\bar{\nu}_e}=  
\langle E_{\bar{\nu}_e}\rangle /3.99)$, 
$\eta_{\nu_x}=1.5$ $(T_{\nu_x}=  \langle E_{\nu_x}\rangle /3.45)$,
$L_{\nu_e}= L_{\bar{\nu}_e}$ and $L_{\nu_e}=1.5 L_{\nu_x}$. 
We fix $\sin^22\theta_{12} = 0.81$, 
$\Delta m^2_{21} = 5.6 \times 10^{-5}$ eV$^2$
and $|\Delta m^2_{31}| = 3 \times 10^{-3}$ eV$^2$, 
since variation of these parameters within their future bounds has very little
effect on the analysis. We generate $\nu_e$ and $\bar{\nu}_e$ 
spectra~\cite{lunardini} and then simulate data by
choosing a point from a Gaussian distribution centered
at the expectation for the bin and of width equal to its square root.
In all, there are four spectra;
one for SK, 
one for the light water tank at SNO, one for processes
 (\ref{neo}) and (\ref{snod}), and one for process 
(\ref{nebard}). The SK spectrum is simulated with 18 bins and the SNO spectra
have 13 bins each. 
We simulate four datasets, two for each type of hierarchy and for
two values of $\sin^22\theta_{13}$ that correspond to adiabatic ($P_H=0$) and 
non-adiabatic ($P_H=1$) oscillation transitions.
We perform a $\chi^2$-analysis, freely
varying $E_b$, $\langle E_{\nu_e} \rangle$, $\langle E_{\bar{\nu}_e}\rangle$,
 $\langle E_{\nu_x}\rangle$, and
$\sin^22\theta_{13}$ to find the $90$\% 
($\Delta \chi^2 < 7.78$)
and $99\%$ C.L. ($\Delta \chi^2 < 13.3$)
allowed regions in  $E_b$, $\langle E_{\bar{\nu}_e}\rangle$,
 $\langle E_{\nu_x}\rangle$ and $\sin^22\theta_{13}$. Although we scan in 
$\langle E_{\nu_e}\rangle$, we do not count it as a free 
parameter since we do not attempt to determine it (knowing a priori of the
limited sensitivity to this parameter).
We
 allow the ratio $L_{\nu_e}/L_{\nu_x}$ to vary
between 1 and 2 to accommodate both perfect equipartitioning and large
departures from it. We fix $L_{\nu_e}=L_{\bar{\nu}_e}$.

Figure~\ref{fig:1reg} shows the results of this fit for the normal hierarchy.
The left-hand
and right-hand panels correspond to data simulated at 
$\sin^22\theta_{13} = 10^{-5}$ (for which $P_H=1$)
and $\sin^22\theta_{13} = 10^{-2}$ (for which $P_H=0$), respectively.  
In either case, we see
that the supernova parameters can be 
determined with high precision, but that $\sin^22\theta_{13}$ is 
unconstrained. 
Since the overall normalization of the neutrino fluxes depends critically on
$E_b$, it is determined with good accuracy. 
The values of $T_{\bar{\nu}_e}$ and $T_{\nu_x}$ are also determined 
precisely since these parameters control the
$\bar{\nu}_e$ spectral distortion (which is independent of
$\sin^22\theta_{13}$) obtained from 
thousands of events at SK and hundreds more at 
SNO. The experiments are not sensitive to $\sin^22\theta_{13}$ because
the $\nu_e$-d events at SNO and 
$\nu_e$-O events at SK and SNO are statistically insufficient.

Figure~\ref{fig:inv} shows the results 
for the inverted hierarchy.
Again, the left-hand and right-hand columns correspond to data
simulated at  $\sin^22\theta_{13} = 10^{-5}$ and 
$\sin^22\theta_{13} = 10^{-2}$, respectively. In the case of non-adiabatic 
transitions, an upper bound on $\sin^22\theta_{13}$ can be placed.
For adiabatic transitions, $\sin^22\theta_{13}$ and
$\vev{E_{\bar{\nu}_e}}$ 
cannot be simultaneously bounded if both are left free. 
When we restrict $\langle E_{\bar{\nu}_e}\rangle$ to lie
between 10.5 MeV and 19 MeV and  
$\langle E_{\nu_x}\rangle/\vev{E_{\bar{\nu}_e}}$ to be larger than
about 0.7 (recall that $\langle E_{\nu_x}\rangle/\vev{E_{\bar{\nu}_e}}$ 
is expected to be larger 
than unity), a lower bound on   
$\sin^22\theta_{13}$ is obtained. 
 For values of 
$\sin^22\theta_{13}$ between $10^{-4}$ and $10^{-3}$, an upper
or lower bound can  be placed on  $\sin^22\theta_{13}$ depending on whether 
it is closer to $10^{-4}$ or to $10^{-3}$.
In the inverted hierarchy, we have a lesser 
sensitivity to $T_{\bar{\nu}_e}$ and $T_{\nu_x}$ since the $\bar{\nu}_e$ flux
is also
sensitive to $\sin^22\theta_{13}$ leading to competition between 
these parameters. Bounds on $\sin^22\theta_{13}$ can be placed because 
the $\bar{\nu}_e$ spectrum is more sensitive to $\sin^22\theta_{13}$
in the inverted hierarchy and the $\bar{\nu}_e$ signal at SK is huge.
 
Although the regions in Figs.~\ref{fig:1reg} and~\ref{fig:inv} are
calculated assuming that the neutrinos detected at SK crossed both the 
mantle and core of
the earth ($\cos \theta_Z=-0.93$), and those at SNO crossed the mantle only 
($\cos \theta_Z=-0.1$), we have established that
the bounds placed are
largely independent of the supernova's zenith angles at the two experiments.

If the mass hierarchy is unknown at the time 
of a supernova signal, it can be deduced 
provided $\sin^22\theta_{13} \gsim 10^{-3}$~\cite{dighe,minakata}, 
and the hierarchy is inverted. 
For values of $\sin^22\theta_{13}$ smaller than $\approx 10^{-4}$, 
$P_H$ is not close to zero and the
survival probabilities are similar for the two hierarchies rendering
them indistinguishable~\cite{dighe}. In the case of a normal 
hierarchy, we see from Fig.~\ref{fig:1reg} that $\sin^22\theta_{13}$ 
is unconstrained even for adiabatic transitions, thereby indicating a
lack of discriminatory power between
$P_H=0$ and $P_H=1$ or equivalently between the mass hierarchies. On the
other hand, for the inverted hierarchy and adiabatic transitions, a lower
bound on $\sin^22\theta_{13}$ can be placed which in turn means that 
the inverted hierarchy can be selected over the normal hierarchy. 
%Again, this is because of the large number of ${\bar{\nu}_e}$ events 
%expected at SK, and the resulting high sensitivity to the inverted hierarchy.

\vskip 0.1in
\noindent
\underline{Future prospects}: 
The next generation of proton decay experiments such as 
Hyper-Kamiokande (HK)~\cite{hyperk} 
and UNO~\cite{uno} are expected to offer 
a new level of sensitivity to the physics of
supernovae and neutrino mixing. We consider the proposed 1 Mton 
HK detector. With no specific information about the detector, we treat it
as a scaled-up version of SK. 
We assume a fiducial volume for supernova neutrinos of 890 kton, 
which is consistent
with the fiducial volume to total volume ratio expected for 
the proposed UNO detector~\cite{uno1}. 

We find that our qualitative 
conclusions for
SK and SNO continue to hold for HK. The quantitative differences are easily
anticipated as a result of its larger volume. The supernova parameters 
can be determined with greater accuracy although $T_{\nu_e}$ will remain 
unknown. In the case of the inverted hierarchy and 
adiabatic transitions, while 
$T_{\nu_x}$ can be determined without theoretical prejudice,
a plausible window has to be chosen for 
$T_{\bar{\nu}_e}$ to constrain $\sin^22\theta_{13}$. 
Also, tighter upper or lower 
bounds can be placed on $\sin^22\theta_{13}$. We emphasize that in the
case of a normal hierarchy, both $\sin^22\theta_{13}$ and the hierarchy 
remain unknown.

\vskip 0.1in
\noindent
\underline{Summary}: 
We have considered what information can be extracted from neutrinos detected
at SNO and SK from a galactic supernova. 
The information they carry is of
major importance in understanding the astrophysics of supernovae. 
The binding energy released in the supernova and the 
temperatures of the non-electron neutrinos expelled may be determined 
with good
precision for most values of $\sin^22\theta_{13}$.
Bounds on $\sin^22\theta_{13}$ can be placed
if the neutrino mass hierarchy is inverted.  
In this case the hierarchy can be determined 
if $\sin^22\theta_{13}\gsim 10^{-3}$.

The above conclusions apply to Hyper-Kamiokande as well.

\vspace{0.1in}
{\it Acknowledgments}:  
We thank E. Beier, G.G. Raffelt, 
E. Kearns, R. Robertson and C. Walter for discussions and
the referees for helpful suggestions.
This research was supported by the U.S.~DOE  
under Grants No.~DE-FG02-95ER40896 and No.~DE-FG02-91ER40676  
and by the WARF.
%\vspace*{-.2in}  

\end{document}